# Modeling COVID-19 vaccine-induced immunological memory development and its links to antibody level and infectiousness

Xin Gao[1,2,#,*], Jianwei Li[1,2,#], Dianjie Li[1,2]

1. School of Physics, Peking University, Beijing, China

2. Center of Quantitative Biology, Peking University, Beijing, China

# These authors contributed equally to this work.

* Corresponding author. Email: ukp@pku.edu.cn

## Abstract

COVID-19 vaccines have proven to be effective against SARS-CoV-2 infection. However, the dynamics of vaccine-induced immunological memory development and neutralizing antibodies generation are not fully understood, limiting vaccine development and vaccination regimen determination. Herein, we constructed a mathematical model to characterize the vaccine-induced immune response based on fitting the viral infection and vaccination datasets. With the example of CoronaVac, we revealed the association between vaccine-induced immunological memory development and neutralizing antibody levels. The establishment of the intact immunological memory requires more than 6 months after the first and second doses, after that a booster shot can induce high levels neutralizing antibodies. By introducing the maximum viral load and recovery time after viral infection, we quantitatively studied the protective effect of vaccines against viral infection. Accordingly, we optimized the vaccination regimen, including dose and vaccination timing, and predicted the effect of the fourth dose. Last, by combining the viral transmission model, we showed the suppression of virus transmission by vaccination, which may be instructive for the development of public health policies.

# Introduction

SARS-COV-2 and its variant of concerns (VOCs) are highly contagious pathogens (*1*), cumulatively resulting in over 340 million infections and more than 5.5 million deaths worldwide by the end of February 2022. Vaccination, as an effective and promising way against SARS-COV-2, reduces the spread of COVID-19 and risk of severe illness significantly (*2*). Clinical studies have demonstrated the safety and efficacy of the vaccines, including inactivated vaccines (such as CoronaVac and Sinopharm), mRNA vaccines (such as Moderna and Pfizer-BioNTech) and other types of COVID-19 vaccines (*3-5*). Therefore, vaccines and booster shots have recognized as an important measure for the prevention of COVID-19 (*6*). However, the dynamic mechanism underlying the protection of vaccines against SARS-COV-2 infection has not been fully elucidated.

Recently, dynamics models have been widely used to study the transmission and infection processes of COVID-19 (*7-10*), helping us to quantitatively analyze the spread of viruses and to establish corresponding public safety policies. In the early stages of COVID-19 epidemic, a study pointed out that COVID-19 would cause over 500,000 deaths in the UK if without government intervention (*8*), which provided the rationale for the need for an epidemic prevention policy in the UK. In addition, another study (*7*) demonstrated that Wuhan closure reduced the number of infections by 96% using SAPHIRE dynamic model, a model combined both infectious properties of COVID-19 and the population mobility. In addition to studying population-level dynamical models, models of within-host viral infection can quantitatively simulate the processes of viral proliferation and clearance, giving a basis for evaluating the medication strategies (*10*). Therefore, dynamics modeling is a promising approach to deepen our understanding of the host immune response induced by vaccines.

Currently, more than 30 approved vaccines are in clinical use and 178 vaccine candidates are in development worldwide (covid19.trackvaccines.org). CoronaVac, as an example, is an inactivated vaccine against COVID-19, containing inactivated SARS-COV-2 virus. The previous study has proved that the first and second doses of CoronaVac vaccination were well tolerated and had moderate immunogenicity in individuals aged 18-59 years (*3*). Dose comparisons revealed that the 6 ug dose of CoronaVac generated slightly higher geometric mean titers (GMTs) of neutralizing antibodies than the 3 ug dose of CoronaVac, and the incidence of adverse reactions were close in 3 ug and 6 ug trials (*11*). Further study found that despite a sharp decline in GMTs 8 months after the second dose, the booster shot increases GMTs by tens of times (*11*). Notably, booster shots inoculated 2 months after the second injection resulted in much lower GMTs than the group that inoculated booster shots 8 months later (*11*). These results show that the formation of intact antibody-associated immunological memory may cost several months in recipients after vaccination (*12*). Through dynamical modeling, we can explain this phenomenon and predict GMTs for a possible fourth shot.

Here, based on viral load (VL) and GMTs data after vaccination, we developed a simplified mathematical model to recapitulate the dynamics of immunogenicity and immunological memory after vaccination with CoronaVac. By simulating the maximum viral load (MVL) and recovery time (RT) after infection, we quantitatively demonstrated the protection of CoronaVac against SARS-COV-2 infection. Subsequently, we optimized the dose of CoronaVac and the booster shot regimen. Finally, we demonstrated the influence of vaccination on viral transmission.

# Results

## Model set

To evaluate the vaccine-induced adaptive immune response and immunological memory, we constructed a simplified within-host immune network mainly focusing on humoral immunity. This network consisted of three modules: virus module ($V$, representing SARS-COV-2 virus and inactivated vaccines), humoral effector module ($E$, representing virus-specific plasma cells and virus-specific neutralizing antibodies) and humoral memory module ($M$, mainly representing virus-specific memory B cells). Fig. 1A illustrates the interaction of these three modules. For virus module $V$, SARS-COV-2 virus proliferates after host infection, while inactivated vaccines do not have the ability to proliferate. SARS-COV-2 virus or inactivated vaccines provide virus-associated antigens stimulation and activate host immunity. In response to antigen stimulation, antigen-specific B cells proliferate and differentiate into plasma cells and memory B cells, and plasma cells secrete neutralizing antibodies to clear the virus. This process, in the model, is described by the activation of $V$ on the proliferation of $E$, the inhibition of $E$ on $V$, and the promotion of $E$ on $M$. Memory B cells can sustain survival in the absence of antigens by replenishing their numbers through homeostatically replicating (*13*). Once memory B cells encounters similar antigen, they will differentiate into effector module, producing a rapid response to specific antigens. This process is be described by the self-proliferation of $M$ and the conversion to $E$ in response to antigenic stimulation.

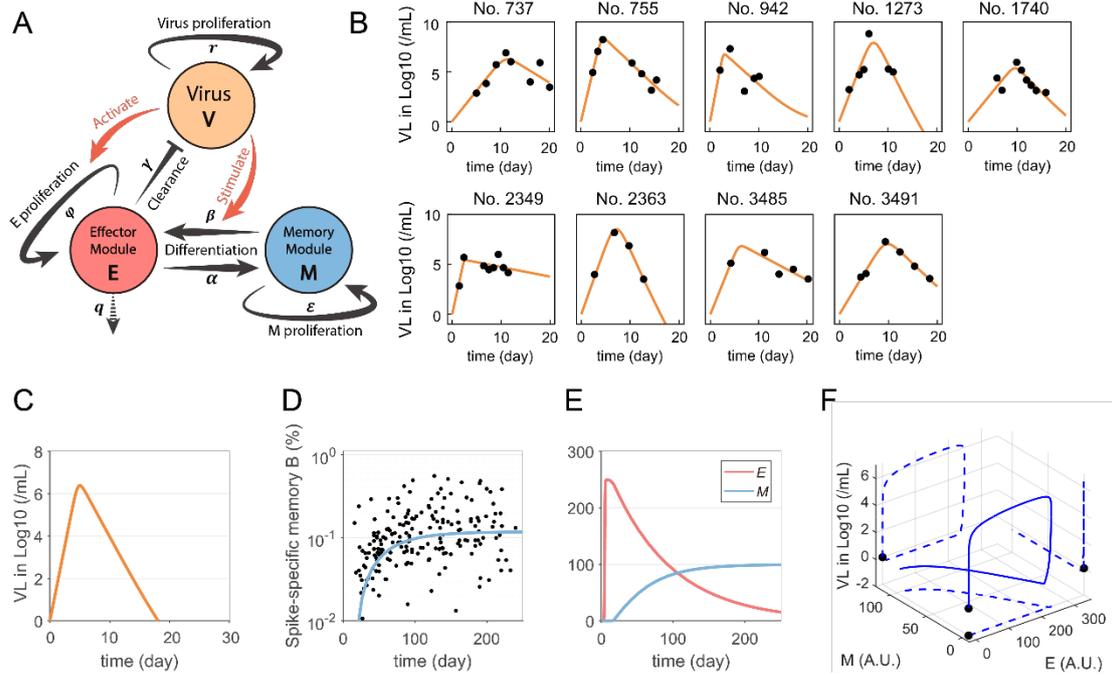

**Fig. 1** Dynamics of viral infection. **A** Schematic diagram of the simplified virus-immune interaction network, focusing on humoral immunity. The effector module ($E$) mainly represents B and plasma cells, and the memory module ($M$) mainly represents memory B cells. Effector module proliferates and differentiates into memory module under stimulation of virus-associated antigens and capable of virus clearance. Memory module self-proliferates and sustain survival even in the absence of antigens and rapidly differentiate into effector module once stimulated by antigens. **B** Results of our model fitting viral load (VL) data from nine individuals in the NBA study of Kissler et al (*10, 14*). **C** Model-predicted mean dynamics process of VL growth and clearance after virus infection. **D** Results of our model fitting spike-specific memory B data from the study of Dan et al (*15*) with the simulation result. In the model, the environmental capacity of $M$ is 100, thus we fit the scaling rate $S_1$=843.9 and use $M/S_1$ to fit the data. **E** Demonstration of effector (red line) and memory (blue line) modules after virus infection. **F** Evolutionary trajectory of viral infection is plotted in the three-dimensional $V$-$E$-$M$ space. The black dots denote the stable-state before infection. In the meantime, we give the projection of the 3-D evolutionary trajectory by dotted lines on 2-D phase plane of $E$-$M$, $E$-$V$ and $M$-$V$ respectively.

Based on the network in Fig. 1A, we build a 3-variable ordinary differential equations model to describe the dynamic process of immune response against SARS-CoV-2 virus as follows:

$$\frac{dV}{dt} = rV - \gamma EV \tag{1}$$

$$\frac{dE}{dt} = \frac{V}{V+V_1}(k+\varphi E)\left(1-\frac{E}{E_m}\right) + \beta VM - qE \tag{2}$$

$$\frac{dM}{dt} = (\alpha E + \varepsilon M)\left(1-\frac{M}{M_m}\right) - \beta VM \tag{3}$$

Where, $V$ represents the VL with a proliferation rate $r$, while proliferation rate equals to 0 when describing inactivated vaccines. $E$ represents the effector module and capable of clearing virus at a rate $\gamma$. The first term of equation (2) describes the proliferation of $E$ under viral antigen stimulation. Proliferation of $E$ is described using a modified logistic growth with the carrying capacity $E_m$; Front coefficient $V/(V+V_1)$ describes the stimulation of virus-associated antigens with saturation effects; $\varphi$ was the proliferation rate while $k$ was a basic growth rate to avoid $E=0$ being a fixed point when $V \neq 0$. The second term of equation (2) describes the activation of memory module $M$ into $E$ at a rate $\beta$ under antigen stimulation, and the third term describes the process of natural apoptosis at a rate $q$. $E$ differentiates into $M$ at the rate $\alpha$. while $M$ self-proliferates with proliferation rate $\varepsilon$ and carrying capacity $M_m$, independent of the antigens. The fitted values of the parameters are list in Table 1.

Table 1. Estimated parameter values in our model.

| Parameters | Descriptions | Values | SD |
|---|---|---|---|
| $r$ | Proliferation rate of virus module | 3.3 day$^{-1}$ | 1.6 |
| $\gamma$ | Clearance of virus module by effector module | 0.018 A.U.$^{-1}$day$^{-1}$ | 0.00065 |
| $V_1$ | Hill constant for effector module activation by virus module | 10000 /mL | Est. |

| $k$ | Basic growth rate of effector module | 1 A.U. day$^{-1}$ | Est. |
|---|---|---|---|
| $\varphi$ | Proliferation rate of effector module | 3.6 day$^{-1}$ | 3.4 |
| $E_m$ | Carrying capacity of effector module | 250 A.U. | Est. |
| $q$ | Natural apoptosis rate of effector module | 0.018 day$^{-1}$ | 0.0016 |
| $\beta$ | Differentiation rate of memory module into effector module due to the stimulation of virus | 0.077 day$^{-1}$ mL | 0.0020 |
| $\alpha$ | Differentiation rate of effector module into memory module | 0.0070 day$^{-1}$ | 0.0062 |
| $\varepsilon$ | Proliferation rate of memory module | 0.020 day$^{-1}$ | 0.065 |
| $M_m$ | Carrying capacity of memory module | 100 A.U. | Est. |
| $S_1$ | Scaling rate of Memory B cells to $M$ | 844 A.U. | 326 |
| $S_2$ | Scaling rate of GMTs to $E$ | 0.42 | 0.0029 |
| $eVL$ | The equivalent VL in CoronaVac | 117 copies/(mL*ug) | 1.6 |

## Dynamic of virus infection

We first investigated the dynamics of viral infection. We adopted the NBA dataset, from the study of Ke et al. and Kissler et al (*10, 14*). The NBA dataset contains VL data from early stage of infections, and early stage data allow us to fit the proliferation process of virus more accurately. We used our model to fit the NBA dataset (see Methods). The results (Fig. 1B) shows that the average viral proliferation rate $r$ for the nine individuals is 3.3 day$^{-1}$ (SD: 1.6); the viral clearance rate $\gamma$ is 0.018 AU$^{-1}$day$^{-1}$ (SD: 0.00065); the proliferation rate of the effector module $\varphi$ is 3.6 day$^{-1}$ (SD: 3.4). We set the proliferation rate of the effector module under antigenic stimulation $k$ to be 1 AU$^{-1}$day$^{-1}$. We assumed carrying capacity of $E$ was $E_m$ to ensure that the virus growth period was shorter than the virus clearance period. $V_1$ is the Hill coefficient for $E$ proliferation induced by antigens, which we estimated to be 10$^4$ /mL (We discuss the value of $V_1$ in Supplementary information 1). The mean recovery time for patients with mild viral infections is around 15 days (95% CI, 13-18) (*16*), which is close to our simulations (13.9 days, Fig. 1C).

In addition to the dynamics of the SARS-CoV-2 infection, we also investigated how the virus or vaccine activates immune memory infection. Immune memory, which provides long-term protection against viruses, consists mainly of memory CD4+ T cells, memory CD8+ T cells, antibodies and memory B cells (*17, 18*), and demonstrate various dynamics (*19*). We focus on the formation of humoral immune memory, especially memory B cells. Dan et al. found that SARS-CoV-2 spike specific memory B cells increase gradually after infection, peaking at 8 months, and will persist for a long time (*15*). We introduced the data of memory B cells and fitted it by our model (Fig. 1D). We assumed that the environmental carrying capacity of memory module $M_m$ to be 100. The fitted proliferation rate of memory module $\varepsilon$ is 0.020 day$^{-1}$ (SD: 0.065) and the fitted differentiation rate of $E$ into $M$ $\alpha$ is 0.0070 day$^{-1}$ (SD: 0.0062). Fig. 1E-F demonstrates the immune response after viral infection, in which the viral clearance takes 13.9 days, while the intact humoral immune memory takes several months to form. Notably, the generation of specific immune response and immune memory differ significantly on the time scale.

## Dynamic of vaccination

Based on the fitted parameters of viral clearance and immune memory formation, we further investigated the vaccine-induced immune response. Considering that inactivated SARS-CoV-2 in CoronaVac is incapable of proliferation, we modify virus proliferation rate $r$ to zero in equation (1). We introduced three doses vaccination data from Zeng et al. (*11*), which contains GMTs measured after 3 ug and 6 ug doses of CoronaVac under four vaccination regimens respectively (Fig. 2A-B). To simulate GMTs with our model, we assumed that the effector module $E$ is proportional to the values of GMT and satisfies $E = S_2 \times GMTs$ (rapid equilibrium assumption, Methods), and we fitted the natural apoptosis rate of effector module $q$ by 0.018 day$^{-1}$ (SD: 0.0016) based on the degradation of GMTs (*11*). We used our model to fit the vaccination data for all regimens simultaneously and best fit result show that the equivalent VL in CoronaVac (*eVL*) is 117 copies/(mL*ug) (SD: 1.6), the differentiation rate of $M$ into $E$, $\beta$, is 0.077 day$^{-1}$ mL (SD: 0.0020), the environmental carrying capacity of $E$, $E_m$, is 250 A.U. (we discuss the influence of $E_m$ in Supplementary Information 1), and $S_2$ is 0.42 (SD: 0.0029).

Based on fitting the model to the vaccine data, we could demonstrate the evolution of $E$ and $M$ under different vaccination regimens (Fig. 2A-B). The results show that GMTs induced by the first and second doses are limited, due to the absence of immunological memory. Also, when the second and third doses were separated by 2 months, immunological memory was not fully formed, therefore, GMTs induced by the third dose did not increase significantly compared to second doses. In contrast, when the second and third doses are separated by 8 months, immunological memory is formed and GMTs induced by the vaccine are the highest, up to tens of times of GMTs before vaccination. Based on the simulation results, we speculated that the first two doses of inactivated vaccination aim at inducing the formation of immunological memory, while the third dose is aimed at inducing the production of neutralizing antibodies after the formation of immunological memory. Currently, the commonly used vaccination regimen of 3 ug CoronaVac is that 21 days between the first and second doses and 6 months between the second and third doses (www.who.int). We simulate the evolution of $E$ and $M$ under this scenario (Fig. 2C-E) and demonstrate that this regimen ensures the preliminary formation of immunological memory and can produce high-level GMTs after the third dose.

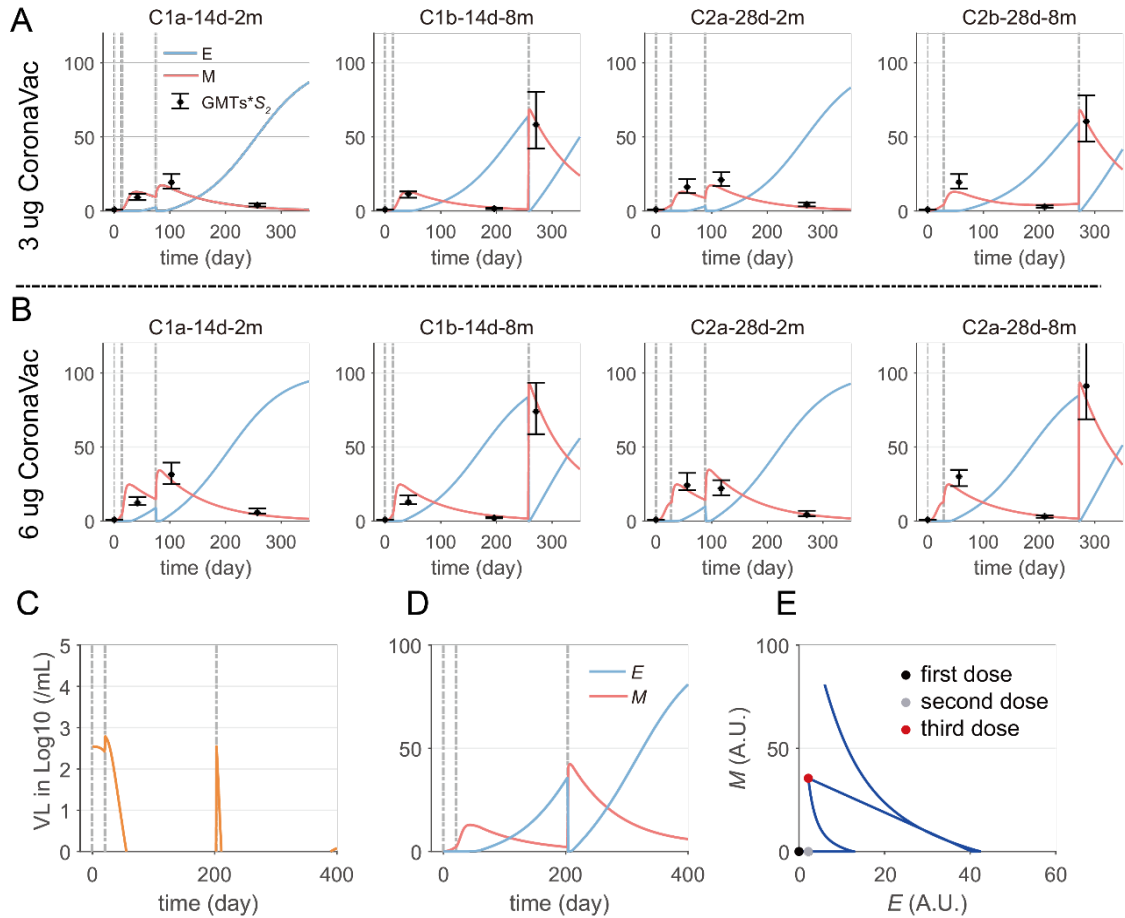

**Fig. 2** Dynamics of vaccination. **A-B** Results of $E$ in our model fitting GMTs under four different vaccination protocols from Zeng et al.(*11*). The gray dashed lines represent the date of vaccination with 3 ug (**A**) or 6 ug (**B**) CoronaVac, the red lines represent $E$, and the blue lines represent $M$. The black dots represent the GMTs divided by 2.5 on the corresponding dates. 'C1a-14d-2m' denote the vaccination protocol that 14 days between the first and second doses and 2 months between the second and third doses. **C-D** demonstrates the VL (**c**), $E$ and $M$ (**D**) after vaccination with 3 ug CoronaVac *in Silico*. The interval between first and second doses is 21 days, and the interval between second and third doses is 182 days, which close to the recommended vaccination protocol. **E** The evolution of $E$ and $M$ in phase diagram after vaccination with 3 ug CoronaVac.

We also found that 6 ug CoronaVac can produce higher level GMTs than 3

ug CoronaVac after both the first and second doses or the third dose and no evidence shows higher adverse effects of the 6 ug CoronaVac than 3 ug. The effect of vaccine dose on GMTs will be discussed in a subsequent section.

## Protection of vaccine against viral infection

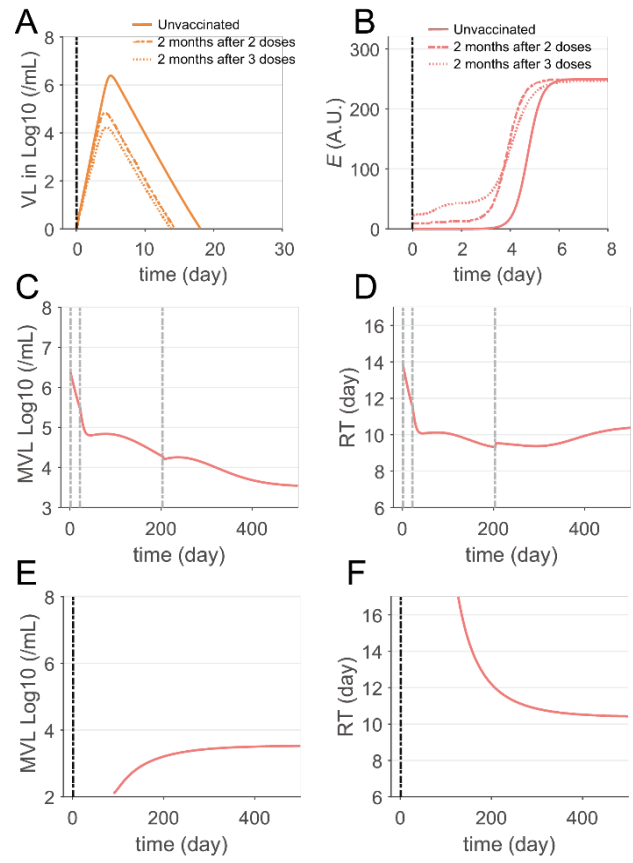

**Fig. 3** Maximal viral load (MVL) and recovery time (RT) of SARS-CoV-2 infection after vaccination. **A** VL after infection with or without vaccination. We compare VL for unvaccinated infections, infections two months after two doses of CoronaVac, and infections two months after three doses of CoronaVac. Black dotted line denotes the day of virus infection. **B** We compare the evolution of effector module $E$ for unvaccinated infections, infections two months after two doses of CoronaVac, and infections two months after three doses of CoronaVac. **C-D** shows the MVL (**C**) and RT (**D**) after infection during the course of a three-dose vaccination. Gray dotted line denotes the day of vaccination. **E-F** For comparison, we give the MVL (**E**) and RT (**F**) after SARS-CoV-2 reinfection. Black

dotted line denotes the day of the first infection.

Simulations and clinical results both have proved that intact vaccination with CoronaVac leads to the production of SARS-CoV-2-specific neutralizing antibodies (*11*). To quantitatively assess the protection of CoronaVac against the infection, we define two indicators here, maximal viral load (MVL) and recovery time (RT, from the moment of infection to the moment that VL below 2). For the unvaccinated population, the fit yielded an average MVL of 6.4 log 10/mL and an average recovery time of 13.9 days. Our result shows that infection two months after two doses CoronaVac, the MVL decreases to 4.8 log10/mL and RT decreases to 10.2 days. Furthermore, an intact three-dose vaccination resulted in further reductions in MVL and RT (4.4 log10/mL and 9.5 days). Vaccination allows the effector immune module to response quickly once exposure to virus, thereby reducing MVL and RT following infection. Fig. 3B demonstrates the response of the effector immune module to infection in unvaccinated, two months after two doses and two months after three doses. It takes about 4 days for the $E$ module of unvaccinated population to proliferate after the viral infection, while an intact three-dose vaccination reduces the response time of $E$ after infection.

Then, we evaluated the changes in MVL (Fig. 3C) and RT (Fig. 3D) for each day after vaccination, if infection occurred. We can observe a significant decrease in MVL and RT after the first and second doses of vaccination and a further decrease in MVL and RT over time. The protection of vaccines against viral infection peaks after the third dose, MVL stabilizes at 3.5 log10/mL and RT stabilizes at 10.4 d. For comparison, we also give the MVL and RT for viral reinfection (Fig. 3E-F, the black line represents the date of the first infection). The results show that the MVL for reinfection is low after recovery from the first infection, indicating that individual is well protected against reinfection. Moreover, the MVL and RT were similar between those who received three complete doses of vaccine and those who were infected with the virus six months later.

## Optimization of vaccination strategy

In a previous three-dose CoronaVac study (*11*), researchers measured GMTs after vaccinated with 3 ug and 6 ug CoronaVac under multiple vaccination regimens. Based on the GMTs data, mathematical model can help us optimize vaccine doses and vaccination regimens for better immunization and lower cost (*20*).

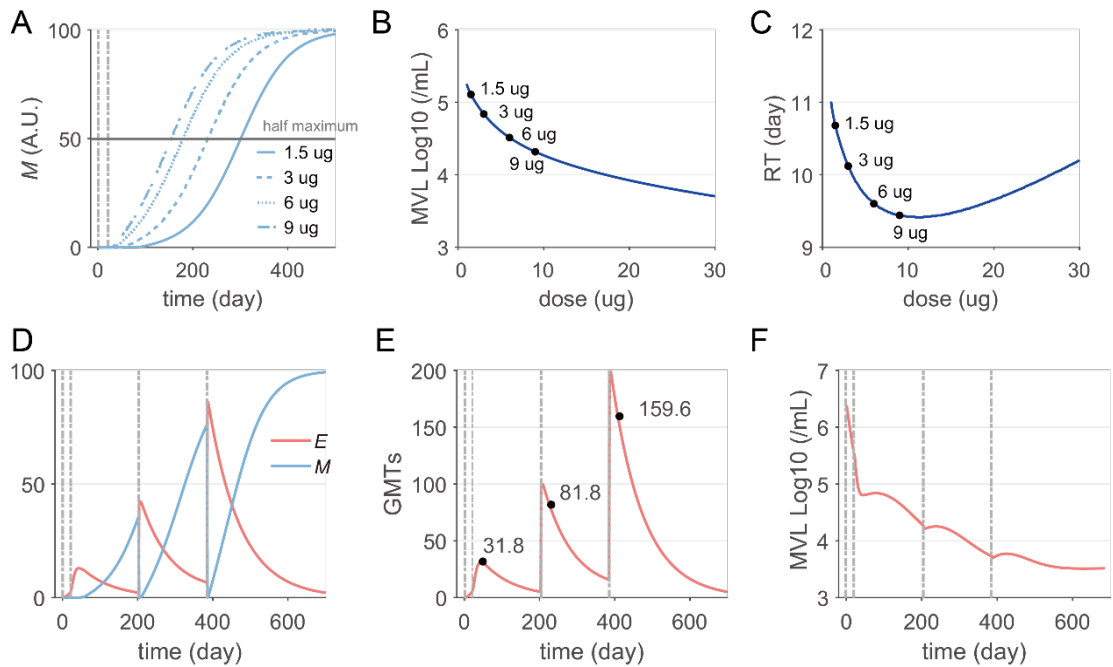

**Fig. 4** Optimization of vaccine doses and strategies. **A** The simulated growth curves of memory module $M$ induced by the first and second shots with 1.5 ug, 3 ug, 6 ug and 9 ug dose CoronaVac. The gray dotted lines denote the date of vaccination. **B-C** The MVL (**B**) and RT (**C**) for infection 2 months after the first and second shots with different doses of CoronaVac. **D** demonstrates the value of $E$ and $M$ module induced by the fourth shot of 3 ug CoronaVac 6 months after the third dose (3 ug CoronaVac, 21 days between the first and second doses and 6 months between second and third doses). **E** demonstrates the GMTs induced by the fourth shot 6 months after the third dose (3 ug CoronaVac, 21 days

between the first and second doses and 6 months between second and third doses). **F** shows the MVL after infection during the course of a three-dose vaccination and the fourth shot.

The first and second doses of vaccines are intended to induce immunological memory formation and can only produce low titers of neutralizing antibodies. Therefore, we investigate the influence of the first and second vaccine doses on immunological memory formation. We simulated the growth curve of memory module $M$ vaccinated with 1.5 ug, 3 ug, 6 ug, and 9 ug CoronaVac (Fig. 4A) and found that the time to half-maximum of $M$ under 1.5 ug, 3 ug, 6 ug, and 9 ug dose of CoronaVac is 301 d, 232 d, 179 d, and 155 d respectively. Simulation of infection two months after 2 doses of the vaccine demonstrates that MVL and RT decrease and gradually saturate with increasing vaccine doses, with both 6 ug and 9 ug CoronaVac leading to low MVL and RT (6 ug: MVL=4.5 log10/mL, RT=9.6 d; 9 ug: MVL=4.3 log10/mL, RT=9.4 d). Furthermore, the side effects caused by the vaccine doses are also an important indicator, and clinical trials have not found more side effects with 6 ug than with 3 ug CoronaVac, while 9 ug CoronaVac has not been clinically validated. Considering the growth curve of *M*, MVL and RT under four doses, we think that for the unvaccinated population, the first and second doses with 6 ug may be better than 3 ug in terms to promote the full formation of immunological memory, but this inference still needs to be verified in more clinical trials.

Besides the vaccine dose, we further try to optimize the vaccination regimen. Considering that many people have completed three doses of the vaccine, we investigate the necessity of a fourth dose to maintain immune efficacy. We simulated the growth curves of $E$ and $M$ with a four-dose vaccination (Fig. 4D, 3 ug CoronaVac, 21 days between the first and second doses, 6 months between second and third doses and a fourth shot 6 months after the third dose) The fourth dose leads to a rapid increase in neutralizing antibody levels, predicting GMTs of 159.6 at 14 days after the fourth dose, higher than the 81.8 after the third vaccine dose (Fig. 4E). The intact immunological memory formation, however, also suggests that the continuation of the fifth dose of CoronaVac would produce GMTs close to those of the fourth dose. Similarly, we evaluate the immune efficacy of the fourth vaccine dose based on the MVL after infection with the virus. The results show that the fourth dose caused a slight decrease in MVL, in conclusion, the fourth dose of the vaccine can increase GMTs to saturation levels but does not have a significant improvement on resistance to viral infection than those who have received three doses of the vaccine.

## Vaccination can prevent viral transmission

SARS-CoV-2 viruses can transmit by a contact from infected person to a recipient, and models of viral transmission have been extensively studied (*21, 22*). Here, we investigated the influence of vaccination on the probability of SARS-CoV-2 transmission. The transmission probability is defined as the probability that a typical exposure of an infected person in a short time $\tau$ leads to infection of a recipient. Ke et al. have proved that the probability of transmission can be written as a function of the VL correlation by a probabilistic model (Fig. 5A, we choose the saturation model here and discuss the power-law model in Supplementary Information 3).

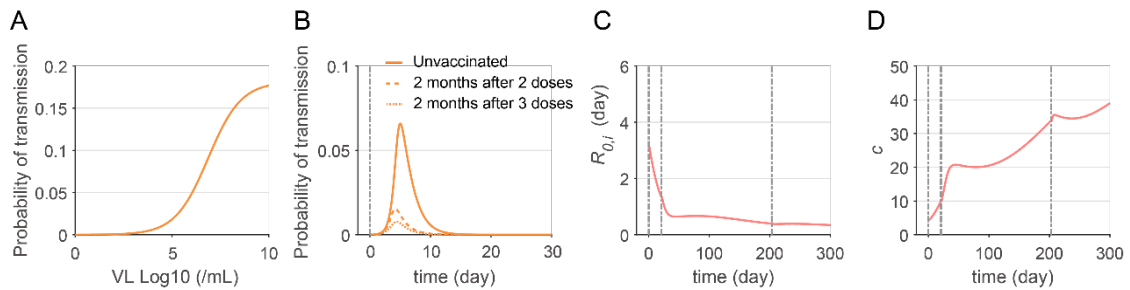

**Fig. 5** Vaccination restricts viral transmission. **A** Saturation model of Jaafar et al. give the probability of transmission for a contact with an individual of different VL. We also investigated the influence of vaccination to the probability of transmission by power-law model (*23*) in Supplementary Information 3. **B** Probability of transmission after infection with or without vaccination. We compared VL for unvaccinated infections, infections two months after two doses of CoronaVac, and infections two months after three doses of CoronaVac. Black dotted line denotes the infection date. **C** shows the basic reproduction number $R_{0,i}$ for individual *i* after infection during the course of a three-dose vaccination. Gray dotted lines denote the vaccination date. **D** demonstrates the maximal contact number *c* to make $R_0$ less than 1 during the course of a three-dose vaccination.

Based on the saturation model describing probability of transmission, we can calculate the probability of transmission *p* after infection by the saturation model (*10*) (see Methods) as:

$$p(t) = 1 - e^{-\theta \frac{V(t)^h}{V(t)^h + K_m^h}}.$$

We demonstrated the probability of transmission after infection in unvaccinated population, vaccinated with 2 doses population and vaccinated with 3 doses population (Fig. 5B). We find that 2-dose vaccination of mutant-specific vaccine reduced the probability of transmission by 77% (from 0.066 to 0.015), while a full 3-dose vaccination reduced the probability of transmission by 88% (from 0.066 to 0.008).

Besides the probability of transmission, $R_0$, the basic reproduction number, is a better-known indicator to measure viral transmission in a population. Social policy, such as wearing face masks, can effectively reduce the value of $R_0$. Similarly, vaccinating a population of mutant-specific vaccine can also reduce $R_0$ (Fig. 5C). The reproduction number for individual i, $R_{0,i}$, can be written as:

$$R_{0,i} = c \int_0^\infty p_i(t)dt,$$

where $c$ is the contact numbers per day and $p_i$ is the probability of transmission for individual i. When $R_0$ of a population is less than 1, the illness will gradually stop spreading. Thus, without considering the differences in the timing of vaccination in a population, we can calculate the maximally allowed contact numbers $c$ per day after vaccination for $R_0$< 1 (Fig. 5D). Considering the average contact number $R_0$ in the EU is 13.4 (*24*), strict social policies are necessary until two months after two doses of vaccines to reduce $c$ and make $R_0$ < 1 in principle.

## Discussion

In this work, we built a mathematical model to quantitatively describe the immune response induced by CoronaVac vaccine and the protection of the CoronaVac against SARS-CoV-2 infection. Through fitting VL data from the unvaccinated population (NBA dataset), we obtained parameters of viral proliferation. We then estimated the proliferation parameters of the humoral memory based on the proliferation curve of spike-specific memory B cells in the infected population. Finally, we simultaneously fitted three-dose GMTs data under eight vaccination regimens (two doses, four vaccination methods) to estimate the differentiation rate of memory module and the equivalent VL of CoronaVac and analyze the differences behind different vaccination regimens.

Based on the data fitting, we have the following findings. First, the first two doses of vaccine only induce limited neutralizing antibodies, mainly to activate the immunological memory, and the complete formation of immune memory takes more than six months. Second, after the formation of the immunological memory, a third dose of vaccine (booster shot) can induce a significant increase in neutralizing antibody levels. Furthermore, to quantitatively characterize the protection of vaccination against viral infection, we defined two indicators, the maximal viral load (MVL) and recovery time (RT) of SARS-CoV-2 infection, and found that vaccination can reduce both MVL and RT.

Then, we explored the optimal vaccine strategy and the impact of vaccines on viral transmission by power-law model (23), using the fitted parameters. We demonstrated that, for the unvaccinated population, increasing the doses of the first and second CoronaVac to 6 ug could promote rapid formation of immunological memory. But this optimization still needs more clinical trials to support. Additionally, we illustrated that two months after the last shot, a 2-dose vaccination reduced the probability of transmission by 77%, while a full 3-dose vaccination reduced it by 88%. However, it was only two months after the second vaccination that the probability of transmission dropped to a relatively low level. This implies that strict social control measures are still necessary to control the spread of COVID-19, when the population is just generally vaccinated with the second dose. Our model can predict the levels of GMTs at different vaccine doses and various regimens, which may be a complement to clinical trials.

This model has limitations in some aspects. First, we only considered humoral immunity in our model, while a complete immune system also contains both innate and cellular immunity. A more complex model is needed to describe the full process of viral infection on the immune system. Second, our model does not take the mutants into account. So far, the dominant strain of COVID-19 has become Omicron in most regions, and vaccines against original strain have limited protection to Omicron (25). However, vaccines against Omicron are in development and our model is still informative for the development of Omicron vaccines. Finally, we only fit the dataset of inactivated vaccines, while the mRNA vaccines produce the spike protein, which may last a few weeks, and maintain long-lasting immunogenicity. mRNA vaccine datasets can be simulated by adding mRNA variables to our model, and we will expand our model to adapt more types of vaccines in future work.

Overall, our model provides a framework and predicts antibody levels induced by inactivated vaccines at different vaccine doses and various regimens, which may be a complement to clinical trials. Furthermore, this model may provide guidance for future vaccine developments.

## Methods

### Estimating parameters

In this model, we performed the fitting of the parameters in three steps to avoid crosstalk of parameters. First, considering that immune memory takes mouths to establish, whose time scale is much larger than the infection, we separated memory module when fitting initial infection and the model degenerated into

$$\frac{dV}{dt} = rV - \gamma EV,$$
$$\frac{dE}{dt} = \frac{V}{V+V_1}(k+\varphi E)\left(1 - \frac{E}{E_m}\right) - qE.$$

We obtained the estimated value of $r$, $\gamma$, and $\varphi$ by fitting the initial infection VL data from NBA dataset (*10, 14*). Second, when fitting the post-infection memory B-cell generation process, we neglected the transition from the memory module to the effector module considering the absence of reactivation of antigen. Then, the model becomes

$$\frac{dV}{dt} = rV - \gamma EV,$$
$$\frac{dE}{dt} = \frac{V}{V+V_1}(k+\varphi E)\left(1 - \frac{E}{E_m}\right) - qE,$$
$$\frac{dM}{dt} = (\alpha E + \varepsilon M)\left(1 - \frac{M}{M_m}\right).$$

Similarly, we obtained the estimated of $S_1, \alpha,$ and $\varepsilon$ by fitting. Third, based on the parameters that obtained above, we fitted the vaccine data using the complete model. This fitting process is performed by fitting antibody levels for all vaccination regimens simultaneously, that is, for a certain set of parameters, the effector module dynamics under different vaccination regimens is calculated simultaneously and then compare with the data. This fitting process helped us to estimate $\beta$, $S_2$, and $eVL$.

We adopted the least squares method for the above fitting, using the leastsq package from scipy. The standard deviation of each parameter was estimated as the arithmetic square root of the diagonal elements of the covariance matrix of the fitted results.

## Parameter sensitivity analysis

To better illustrate the impact of parameters on simulation results, we used SALib package for parametric sensitivity analysis. First, we defined the impact (Err) as the deviation from the original trajectory in the space of $V$, $E$, and $M$:

$$\mathrm{Err} = \sum_i \int \left( T_i^{'}(t) - T_i(t) \right)^2 dt / \max(T_i(t))^2$$

Where, $T_i(t)$ is the trajectory of the $i^{th}$ variable and $T_i^{'}(t)$ is the trajectory of the ith variable after changing the parameters. Summation $i$ from 1 to 3 to consider changes in all three dimensions. Also, to make the impact between different variables to be comparable, we adapted the weighted average by using the square of the inverse of the original trajectory's maximum as the weighting factor. In the program implementation, we used the saltelli.sample function (a Saltelli's sampling scheme generator) to sample the values of parameters within the range of [90%,110%]. With the input N=1024, we obtained 24576 samples of parameters set and performed parametric sensitivity analysis by using Sobol scheme. We showed the total-order indexes of both viral infection trajectory and three doses vaccination trajectory in Supplementaty Information 2.

## Probability of transmission

To describe the association between VL and probability of transmission (*p*), we introduced the study of Ke et al. (10) and calculated the probability of transmission for a typical contact at time t by the saturation model as

$$p(t) = 1 - e^{-\theta \frac{V(t)^h}{V(t)^h + K_m^h}},$$

where *V* is the measured VL. The value of $\theta$, $K_m$ and $h$ are 0.2, 8.9*10$^6$ copies/mL and 0.51 respectively (10). Besides the saturation model, the probability of transmission can also be calculated by the power-law function as

$$p(t) = 1 - e^{-\phi V(t)^h},$$

where $\phi$=2.4*10$^{-5}$ and *h*=0.53 (10). We compared the probability of transmission after vaccination calculated by the saturation model and the power-law model in Supplementary Information 3.

## Acknowledgements

We thank Ziheng Zhao and Fangting Li for useful discussion.

## Author contributions

X.G. conceived the idea for the manuscript. X.G. and J.W.L. contributed to the development of the mathematical model, J.W.L fitted the data. X.G. and J.W.L. wrote the manuscript.

## Competing interests

The authors declare that they have no competing interests.

## Data and materials available

There are no original data in this work. Previously published data were used for this study (*10, 11, 14, 15*).